\documentstyle[12pt]{article}
 
\input epsf
\begin{document}

\begin{titlepage}
\null\vspace{-62pt}
 
\pagestyle{empty}

\begin{center}

\rightline{CCNY-HEP-00/6}
\rightline{hep-ph/0011015}
 
\vspace{1.0truein}
 
{\Large \bf Derivative expansion and gauge independence 
       of the false vacuum decay rate
      in various gauges}

\vspace{0.7in}
 
D. METAXAS \\
\vspace{0.5in}

{\it Physics Department\\
City College of the City University of New York\\
NY, NY  10031\\
metaxas@theory.sci.ccny.cuny.edu}\\
 
\vspace{0.5in}
 
\vspace{.5in}
\centerline{\bf Abstract}
 
\baselineskip 18pt
\end{center}

In theories with radiative symmetry breaking, the
calculation of the false vacuum decay rate requires
the inclusion of higher-order terms in the derivative
expansion of the effective action. I show here that,
in the case of 
covariant gauges, the presence of infrared singularities
forbids the consistent calculation by keeping the
lowest-order terms. The situation is remedied, however,
in the case of $R_{\xi}$ gauges.
Using the Nielsen identities
I show that the final result is gauge independent
for generic values of the gauge parameter $v$
that are not anomalously small.

\end{titlepage}
 
\newpage
\pagestyle{plain}
\setcounter{page}{1}
\newpage
 
\section{Introduction}
 
The usual approach to the calculation of the 
false vacuum decay rate $\Gamma$ yields an
expression of the form \cite{coleman}
\begin{equation}
\Gamma = A\, e^{-B}
\end{equation}
where $B$ is the classical action of the bounce solution
of the Euclidean field equations and $A$ is expression
involving functional determinants which is generally
of order one times a characteristic dimensionful quantity
of the theory.

In theories with radiative symmetry breaking, however, this
approach should be modified since the classical action
may not even have a bounce solution.
The obvious generalization would be to use
the derivative expansion of the quantum Euclidean
effective action
\begin{equation}
S_{\rm eff} = \int \, d^4 x
\left[ V_{\rm eff}(\phi) + \frac{1}{2} Z(\phi)
                      (\partial_{\mu}\phi)^2 \,+ \cdots
\right]
\label{Seff}
\end{equation}
combined with  a power series expansion in the 
coupling constants
\begin{equation}
V_{\rm eff}(\phi) = V_{e^4} + V_{e^6} + \cdots
\end{equation}
 \begin{equation}
Z(\phi) = 1 + Z_{e^2} + \cdots
\end{equation}
where from now on  we assume the general case of a gauge
field theory with scalar coupling of order $e^4$,
where the one-loop radiative corrections
generate a symmetry-breaking minimum at
$\phi =\sigma$, in addition to a symmetric minimum
at  $\phi = 0$.

One can then proceed to find a bounce solution
to the equation
\begin{equation}
\Box \phi_{b} = \frac{\partial V_{e^4}}
                     {\partial \phi}
\label{bounce}
\end{equation}
and get a more complete expression for
the false vacuum decay rate \cite{ejw}:
\begin{equation}
 \Gamma = A'  \,e^{-(B_0 +B_1)}
\label{decayrate} 
\end{equation}
where
 \begin{equation}
B_0 = \int \, d^4 x \left[ V_{e^4}(\phi_b)
      + \frac{1}{2} (\partial_{\mu} \phi_b )^2
            \right]
\label{B0} 
\end{equation}
\begin{equation}
B_1 = \int \, d^4 x \left[ V_{e^6}(\phi_b)
       + \frac{1}{2} Z_{e^2}(\phi_b)
                           (\partial_{\mu} \phi_b )^2
                    \right]
\label{B1}
\end{equation}

The bounce configuration, being a solution of
Eq.(\ref{bounce}), has a spatial extend of order
$(e^2 \sigma)^{-1}$, and accordingly,    
$B_0$ is of order $e^{-4}$ and     
$B_1$ is of order $e^{-2}$. 

The prefactor $A'$ is the result of
functional integrations, which includes 
 higher-order terms in the effective action,
and would be given by a dimensionful factor
times a numerical factor which should be of
order one, otherwise the approximations
made in deriving Eq.(\ref{decayrate})
would break down.

The total expression  for the vacuum decay
rate, being a physical quantity, should 
be gauge independent.
Since the leading order terms in
the effective potential are gauge independent,
it is obvious that the bounce solution
and $B_0$ are gauge independent.
The higher-order terms in the effective action,
however, are gauge-dependent. 

It was shown in \cite{metaxas} that in the 
case of scalar electrodynamics,  
with a Lagrangian
\begin{equation}
  {\cal L}=-\frac{1}{4}F_{\mu\nu}^2
    +\frac{1}{2} (\partial_\mu\Phi_1 - e A_\mu \Phi_2)^2+
     \frac{1}{2} (\partial_\mu \Phi_2 + e A_\mu \Phi_1)^2
       - V(\Phi)
        \label{QED}
\end{equation}
where
\begin{equation}
      V(\Phi) =
      \frac{1}{2}m^2 \Phi^2
      + \frac{\lambda}{4!} \Phi^4
\end{equation}
and $\Phi \equiv (\Phi_1^2 +\Phi_2^2)^{1/2}$,
in the
   general class of $R_{\xi}$ gauges, with a
gauge-fixing term
\begin{equation}
{\cal L}_{gf} = - \frac{1}{2\xi}
 ( \partial_{\mu} A^{\mu}  +  ev \Phi_2 )^2
\label{gf}
\end{equation}
the various terms in $B_1$ combine to give
a $\xi$-independent result.
A similar result regarding the 
$\xi$-independence of the bounce effective
action was derived in \cite{baacke}
in the case of the $SU(2)$-Higgs model,
also in the case of $R_{\xi}$ gauges.

Here I will prove the $v$-independence 
of the false vacuum decay rate in the
case of scalar QED in the class of $R_{\xi}$ gauges,
for generic (of order $\sigma$) values of the
gauge parameter $v$.
I will also describe how the calculation
breaks down for values of $v$ that
are anomalously small.

The class of covariant gauges can be formally
obtained from the class of $R_{\xi}$ gauges
by setting the parameter $v$ to zero.
One would therefore expect the results
for the calculation of the false vacuum 
decay rate mentioned above for the $R_{\xi}$ gauges
to carry over to the case
of the covariant gauges.
I will show in this Paper that this is not 
the case, and the situation in covariant gauges
is quite different.
In particular, I will show that higher-order terms
in the derivative expansion of the effective action
turn out to contribute at the same order of
magnitude as the terms appearing in $B_1$, 
causing the entire formalism to break down.
This is equivalent to saying that the 
prefactor $A'$ in Eq.~(\ref{decayrate})
grows larger, of the order of the exponential,
making the  approximations used in deriving
Eq.~(\ref{decayrate}) in \cite{ejw} invalid.
In other words,  $A'$ and $e^{-B1}$ are
of the same order of magnitude,
are both gauge-dependent, and
it is only the combined
expression that should be gauge-independent.
I do not have a proof of this fact
in covariant gauges since I do not 
have a way of calculating the prefactor
analytically.
A possible way to investigate this would be
along the lines of the calculation
in \cite{baacke}.

In view of this fact, and the fact that 
covariant gauges can be formally obtained from
$R_{\xi}$ gauges by setting $v$ to zero,
it becomes interesting to check the 
gauge-independence of the result for $B_1$ 
in $R_{\xi}$ gauges, now with respect to
the other gauge parameter $v$. 
Using a set of Nielsen identities,
similar to the ones used in \cite{metaxas},
I will prove that the result is indeed
$v$-independent, for generic values of
the parameter $v$, generic meaning here
of order $\sigma$.
As $v$ gets smaller, of order $e^2 \sigma$ 
or less, 
higher-order diagrams start to contribute,
and  the whole calculation breaks down.
It seems that $v$ acts as a sort of infrared
regulator that cures the infrared divergences
of the covariant gauges.
The price to pay is that it is an additional
gauge parameter, and gauge-independence
should be checked with respect to $v$
as well.

Similar considerations and results of this paper
would apply to any calculation of the false vacuum
decay rate in theories where the radiative corrections
are important. This includes theories with
radiative symmetry breaking, and field theories
at high temperature. Although the conclusions 
may vary for different models, the considerations 
for gauge independence via the Nielsen identities
are an important check of the consistency of the
calculation, and can be easily carried over
to other models.

There are various calculations of the false
vacuum decay rate in the framework of the
$SU(2)$-Higgs model at finite temperature
\cite{Kr, Bod, Sur}.
The analysis presented here can be extended
to these models. The Nielsen identities
are a special case of Ward identities that can
be derived with sufficient generality \cite{Kobes}
and can be verified algebraically in many cases
\cite{af}.
In particular, the generalization to
finite temperature and non-abelian theories
is straightforward \cite{Kobes},
although the verification may be more involved
and the conclusions may be different
(different models may have more or less
severe divergences).

The organization of this paper is the following:
In Sec.~2 I will describe the general method
for checking the gauge independence
of the false vacuum decay rate using
the Nielsen identities.
In Sec.~3 I show that the formula for the 
calculation of the decay rate breaks down
in the case of covariant gauges because of
higher-order terms in the effective action, and
I show how this
is reflected in the Nielsen identities.
In Sec.~4 I show that in the case
of $R_{\xi}$ gauges the above problems do not exist
for generic (of order $\sigma$) values of the
gauge parameter $v$, and I show that 
the final value for $B_1$ is
$v$-independent.
Sec.~5 contains some concluding remarks.

\section{The Nielsen identities}

\subsection{Basics}

Consider a gauge theory with a set of
fields denoted collectively by $\phi_i$, with
a classical action that is invariant under
the infinitesimal gauge transformations
\begin{equation}
\delta_g \phi_i = \Delta_i \theta
\end{equation}
where $\Delta_i$ are linear operators,
and $\theta$ an abelian gauge parameter
(the results can be extended to non-abelian
gauge theories, but we will not consider that here).

After choosing a gauge-fixing function
$F(\phi_i)$ and introducing Fadeev-Popov
ghosts $\eta$ and $\bar{\eta}$, 
one can proceed to a path-integral quantization
and obtain the quantum effective action, $S_{eff}(\phi_i)$,
which is the generating funtional of the connected,
one-particle irreducible Green's functions.
The effective action, like the individual Green's functions,
is not a  physical quantity, and
depends on the gauge-fixing condition.

One can derive a general formula that
describes the 
change $\Delta S_{eff}$
of the effective action
because of an infinitesimal change
$\Delta F$ of the gauge-fixing condition \cite{Kobes}: 
\begin{equation}
     \Delta S_{\rm eff} = i \int d^4x\, d^4y\,
    \frac{\delta S_{\rm eff}}{\delta\phi_i(y)}
      \langle \Delta_i\eta(y)\, \bar\eta(x)\, \Delta F(x)
          \rangle_{\rm 1PI}
    \label{nielsen1}
\end{equation}
where the subscript $\rm 1PI$ indicates that only
one-particle irreducible graphs are to be
included. In particular, an infinitesimal change $d\xi$ in the
gauge parameter is equivalent to the choice $\Delta F = -(F/2
\xi) d\xi$.  Hence,
\begin{equation}
    \xi \frac{\partial S_{\rm eff}}{\partial\xi}
    = \int d^4y
         \frac{\delta S_{\rm eff}}{\delta\phi_i(y)} H_i[\phi(z),y]
     \label{nielsen2}
\end{equation}
where
\begin{equation}
     H_i[\phi(z),y]  = -\frac{i}{2} \int d^4x
     \langle \Delta_i\eta(y)\,\bar\eta(x)\,  F(x) \rangle_{\rm 1PI} \,\, .
\end{equation}
Eq.~(\ref{nielsen2}) is the Nielsen identity \cite{nielsen},
which can be used
to prove the gauge independence of various physical quantities.

Our interest 
is to study the false vacuum decay rate
given by Eq.~(\ref{decayrate}).
For simplicity we consider the case
where the effective action depends on a single
field $\phi(x)$, and make a derivative expansion
of the single functional $H$ that enters
the Nielsen identity:
\begin{equation}
       H[\phi(x),y] = C(\phi) + D(\phi)(\partial_{\mu}\phi)^2 +\cdots
\label{CD} 
\end{equation}
Inserting this expansion
together with the derivative expansion of
the effective action into Eq.~(\ref{nielsen2}) we get
\begin{eqnarray}
    &&   \xi \frac{\partial}{\partial\xi}
      \int d^4x \left[V_{\rm eff}(\phi)
     + \frac{1}{2} Z(\phi)(\partial_\mu\phi)^2
     + \cdots \right]  \nonumber \\
   &&\qquad\quad =  \int d^4x \Bigg[C(\phi) + D(\phi)(\partial_\mu\phi)^2
       +\cdots \Bigg]
          \nonumber \\
  && \left[ \frac{\partial V_{\rm eff}}{\partial\phi} +\frac{1}{2}
      \frac{\partial Z}{\partial\phi} (\partial_\mu\phi)^2
   - \partial_\mu \left[ Z(\phi)\,\partial_\mu \phi \right] + \cdots
\right] \, .
    \nonumber \\
\end{eqnarray}
 
Since this equation holds for arbitrary $\phi(x)$ the
terms with equal number of derivatives must be  equal.
From the terms with no derivatives 
and from the terms with two derivatives
we get respectively:
\begin{equation}
    \xi \frac{\partial V_{\rm eff}}{\partial\xi} =
     C \frac{\partial V_{\rm eff}}{\partial\phi}
    \label{V}
\end{equation}
\begin{equation}
    \xi \frac{\partial Z}{\partial\xi} =
           C \frac{\partial Z}{\partial\phi}
         + 2 D\frac{\partial V_{\rm eff}}{\partial\phi}
         + 2 Z\frac{\partial C}{\partial\phi} \, .
    \label{Z}
\end{equation}

\subsection{Scalar QED}

Now we specialize to
the case of scalar QED with 
gauge coupling $e$ and scalar self-coupling
$\lambda$ of order $e^4$, that exhibits
radiative symmetry breaking.
We also limit ourselves to the class of
$R_{\xi}$ gauges, since the situation is
different for covariant gauges,
as we will show in the next section.

For the Lagrangian  in Eq.~(\ref{QED})
and the gauge-fixing in Eq.~(\ref{gf}),
the effective action can be obtained
by making a shift
$\Phi_1 \rightarrow \Phi_1 + \phi$
and dropping the linear terms \cite{jackiw}.
The vertices can be read off from the 
remaining terms, and will depend on $\phi$.
The effective $\Phi_1$, $\Phi_2$ and $A_{\mu}$
propagators
are
\begin{equation}
G_1 (k) = \frac{i}{k^2 - m_1^2}
\end{equation}
\begin{equation}
G_2 (k) = \frac{ i(k^2 - \xi e^2 \phi^2}
               {D(k)}
\end{equation}
\begin{eqnarray}
G_{\mu \nu} (k) &=& G_{\mu \nu}^T (k) +
                    G_{\mu \nu}^L (k)
       \nonumber \\
&=& -i \frac{g_{\mu \nu} - \frac{k_{\mu}k_{\nu}}{k^2}}
            {k^2 -e^2 \phi^2}
   - \frac{i(\xi k^2 - e^2 v^2)}{D(k)}
      \frac{k_{\mu}k_{\nu}}{k^2}
\end{eqnarray}
the mixed $A_{\mu} - \Phi_2$ propagator
is 
\begin{equation}
G_{\mu 2} = -\frac{e(\xi \phi + v) k_{\mu}}
                  {D(k)}
 \end{equation}
for momentum flow from $A_{\mu}$ to $\Phi_2$,
and the ghost propagator is
\begin{equation}
G_g = \frac{i}{k^2 + e^2 v \phi}
\end{equation}

In these expressions
\begin{equation}
      D(k) = k^4 - k^2 (  m^2_2  -2e^2 v\phi)
    + e^2 \phi^2 (e^2v^2 + \xi m_2^2 )
    \label{Ddef}
\end{equation}
and the various  masses that appear
are not the tree-level masses that one
would have in theories without 
radiative symmetry breaking:
\begin{eqnarray}
   &&   m^2_1(\phi) = m^2 + \frac\lambda{2} \phi^2
  =  V''(\phi)  \\
   &&   m^2_2(\phi)=  m^2 + \frac\lambda{6} \phi^2
   =  \frac{V'(\phi)}{\phi}  \, .
\end{eqnarray}

Instead, in our case of radiative symmetry breaking,
the fact that
$\lambda$ is of order $e^4$ 
implies that some
multi-loop graphs are of the same order as  graphs with fewer
loops. In particular, the insertions of photon loops along a
scalar propagator must be resummed.
This can be done simply by replacing the  above 
bare masses by dressed masses given by:
\begin{eqnarray}
   &&  m^2_1(\phi) =  V''_{e^4}(\phi)  \\
   &&  m^2_2(\phi) = \frac{V'_{e^4}(\phi)}{\phi} \, .
   \label{m2mass}
\end{eqnarray}
These are the masses that should be 
considered in the above propagators.

The function $C(\phi)$ in the derivative expansion
(\ref{CD}) can be obtained, to lowest
order, from the two graphs with
operator insertions shown in
Fig.~1.
Their sum is 
of order $e^2$ and given by
\begin{eqnarray}
&&     C_{e^2} = -{ie\over 2} \int{ d^4k \over (2\pi)^4}
      {1 \over (k^2 + e^2 v \phi)\, D(k)}
   \left[ e (\xi \phi +v)\, k^2   - ev( k^2 -\xi e^2 \phi^2 ) \right]
            \nonumber \\
&&  \quad \,\,\,\,= -{ie^2 \phi \xi\over 2}
         \int{ d^4k \over (2\pi)^4} {1\over D(k)} \, .
    \label{Ce2}
\end{eqnarray}
 
The next function $D(\phi)$ in the derivative 
expansion (\ref{CD})
is obtained from similar graphs with
two external $\phi$ lines carrying
momenta $p$ and $-p$.
Examination of these graphs shows
that in the case of $R_{\xi}$ gauges
the lowest order  graphs are of order one.
 
The expansion of the effective potential
in power series in the coupling constant
starts with the term of order $e^4$.
The tree-level potential combines
with the transverse photon loop to give
\begin{equation}
 V_{e^4} = \frac{1}{2} m^2 \phi^2 + \frac{\lambda}{4!}\phi^4
     -\frac{3i}{2} \int {d^4k\over (2\pi)^4} \ln (k^2-e^2\phi^2) \, .
\label{Ve4}
\end{equation}

The next term in the power series
for the effective potential is of
order $e^6$ and is given by
\begin{equation}
       V_{e^6} =
          -\frac{i}{2} \int{ d^4k\over (2\pi)^4} \left [\ln D(k) -
     2\ln (k^2 + e^2 v \phi) \right]
    \label{Ve6}
\end{equation}
where the first term comes from the 
one loop graphs with $\Phi_2$, longitudinal
photon, or mixed scalar-photon propagators and the second 
term from the one loop ghost diagram.

The next term in the derivative
expansion of the effective action
(\ref{Seff}) is given by
\begin{equation}
       Z = - i \left.{\partial \over \partial p^2} \sum I_j
        \right|_{p^2=0}   \, .
\end{equation}
where $I_j(p^2)$ are the contributions of the diagrams
shown in Fig.~2 with two external  $\phi$ lines
with momenta $p$ and $-p$.
In Fig.~2 we show only the diagrams
of the lowest order,  which is $e^2$.

\subsection{Gauge independence}

Now we can study the implications of the
identities (\ref{V}) and (\ref{Z}) when we consider the terms
of the same order in the coupling constant.
From (\ref{V}), the terms of order $e^4$ give
\begin{equation}
   \xi \frac{\partial V^{\rm eff}_{e^4}}{\partial\xi} =0  \, .
   \label{Ve4id}
\end{equation}
which is the well-known result that the leading
order term in the effective potential is 
gauge-independent \cite{jackiw}, and it is
obviously satisfied by (\ref{Ve4}).

The terms of order $e^6$ in (\ref{V})
give
\begin{equation}
   \xi \frac{\partial V^{\rm eff}_{e^6}} {\partial\xi}
    =  C_{e^2}\,  \frac{\partial V^{\rm eff}_{e^4}}{\partial\phi}   \, ,
   \label{Ve6id}
\end{equation}
which is the original form of
the Nielsen identity \cite{nielsen}
that was used to prove the gauge independence
of the phenomenon of symmetry breaking
and other physical quantities.

The terms of order $e^2$ in (\ref{Z}) give
\begin{equation}
   \frac{1}{2} \xi\frac{\partial Z_{e^2}} {\partial\xi}
     =   \frac{\partial C_{e^2}}{\partial\phi} \, .
   \label{Ze2id}
\end{equation}
This is a new form of a Nielsen identity
that was derived in \cite{metaxas}
for the case of $R_{\xi}$ gauges.

Now we are ready to check the gauge-independence
of the  decay rate (\ref{decayrate}).
Using the identities
(\ref{Ve6id}) and (\ref{Ze2id})
we have
\begin{eqnarray}
      \xi \frac{\partial B_1}{\partial\xi} &=&
        \xi \frac{\partial}{\partial\xi} \int d^4x \left[
      V^{\rm eff}_{e^6} +
      \frac12 Z_{e^2} (\partial_\mu\phi_b)^2  \right]  \nonumber \\
     &=& \int d^4x \left[
      C_{e^2} \frac{\partial V^{\rm eff}_{e^4}}{\partial \phi}
      + \frac{\partial C_{e^2}}{\partial \phi} (\partial_\mu\phi_b)^2
       \right]   \nonumber \\
     &=& \int d^4x \left[
      C_{e^2} \frac{\partial V^{\rm eff}_{e^4}}{\partial \phi}
        + (\partial_\mu C_{e^2}) (\partial_\mu \phi_b) \right] \nonumber \\
     &=& \int d^4x \, C_{e^2}
    \left[ \frac{\partial V^{\rm eff}_{e^4}}{\partial \phi} - \Box \phi_b
        \right]
\label{proof}
\end{eqnarray}
and 
the last expression vanishes from the 
definition of the bounce.

The identities (\ref{Ve6id}) and (\ref{Ze2id})
where explicitly verified in
\cite{metaxas} for the case of scaler QED
in the class of $R_{\xi}$ gauges,
and the result for the decay rate
was shown to be $\xi$-independent
for generic values of the 
gauge parameter $v$.
In the next section I will show
that these identities break down 
in the case of covariant gauges,
signaling the breakdown of the 
entire derivative expansion employed
in the calculation of the decay rate from
(\ref{decayrate}).
Then in Section 4 I will show that,
in spite of the infrared divergencies
that plague the covariant gauges,
the calculation in $R_{\xi}$ gauges is
well-defined with respect to the other
gauge parameter, $v$, and the final
result for the decay rate is $v$-independent.
As mentioned before, this holds
for generic values of the 
gauge parameter $v$, generic meaning 
of order $\sigma$.

\section{The problems in covariant gauges}

The situation in covariant gauges is quite 
different, and essentially the entire
formalism described before for the
calculation of the false vacuum decay rate
breaks down \cite{stathakis}.
Technically, the problems come from
contributions of diagrams with 
longitudinal photon propagators.
The function that appears in the
denominator of the longitudinal photon propagator
becomes
\begin{equation}
      D(k) = k^4 - k^2 m^2_2
    + e^2 \phi^2 \xi m_2^2 
    \label{Ddefcov}
\end{equation}
and the product of its roots
is of order $e^6$
since $m_2^2$ is of order $e^4$.
The product of the roots of the
corresponding function in $R_{\xi}$ gauges,
(\ref{Ddef}), is of order $e^4$
for generic (of order $\sigma$) values 
of the gauge parameter $v$.

Consider a term of the next order in the
derivative expansion of the bounce effective action,
a term with four derivatives, of the form
\begin{equation}
\int d^4x \, W(\phi) \, (\partial_\mu \phi_b )^2
    \, (\partial_\mu \phi_b )^2 \,\,.
\label{W}
\end{equation}
The function $W(\phi)$ receives contributions
from graphs like the one in Fig.~3
with four external lines carrying momenta
$p$ and $-p$.
It is easy to see that the $\xi$-dependent
contributions
of the terms with longitudinal photons
in this graph are at least of order $e^{-2}$.

Remembering that the bounce has a spatial
extent of order $(e^2 \sigma)^{-1}$, 
we see that the contribution of (\ref{W})
is of order $e^{-2}$, same as the factor 
$B_1$ in (\ref{B1}).
This means that the calculation of the 
false vacuum decay rate from
(\ref{decayrate}) is not self-consistent,
as there are other contributions comparable to $B_1$,
coming from higher order terms in the effective action.
In fact, one can easily see through power-counting
that all higher order terms have comparable 
contributions.

This situation is reflected in the Nielsen
identities.
Consider a graph of the form shown in Fig.~4,
with external momenta $p$ and $-p$,
that contributes to the calculation of the
quantity $D(\phi)$ appearing in (\ref{Z}).
An easy calculation shows that the part
of the graph coming from the longitudinal
photons gives a contribution to  $D(\phi)$ 
of order $e^{-2}$.
Accordingly, instead of (\ref{Ze2id})
we have 
\begin{equation}
   \frac{1}{2} \xi\frac{\partial Z_{e^2}} {\partial\xi}
     =   \frac{\partial C_{e^2}}{\partial\phi} 
        + D \frac{\partial V_{e^4}}{\partial \phi} \,.
   \label{Ze2id2}
\end{equation}
The calculation of (\ref{proof}) does not hold,
and the quantity $B_1$ is gauge dependent.
The combination of $e^{-B_1}$ and the terms
of higher order in the derivative expansion,
like (\ref{W}), should be gauge-independent,
but it does not seem possible to
prove this perturbatively.

These problems do not arise in the case 
of $R_{\xi}$ gauges, where the corresponding
expressions for (\ref{W}) and $D(\phi)$
are of order unity for 
values of $v$ of order $\sigma$ or larger.
It is only for values of $v$ of order
$e^2 \sigma$ or smaller that the 
calculation starts to break down.
It seems that the gauge parameter $v$
acts as a sort of infrared regulator,
but of course the price to pay is that it
is an additional arbitrary gauge parameter,
and one has to check the gauge independence
with respect to $v$ as well.
I will do that in the next section.

\section{Gauge independence in $R_{\xi}$ gauges}

The $\xi$-independence of the decay rate
in $R_{\xi}$ gauges was proven in \cite{metaxas}.
Here I will prove that the expression
$B_1$ is also $v$-independent.
Similar investigations of the gauge independence
of other physical quantities in $R_{\xi}$ gauges 
have been done in \cite{af}.

Starting from the most general form of the
Nielsen identity, Eq.~(\ref{nielsen1}),
one gets the change of the effective action 
with respect to the parameter $v$: 
\begin{equation}
     \frac{\partial S_{\rm eff}}{\partial v}
    = \int d^4y
         \frac{\delta S_{\rm eff}}{\delta\phi(y)} H^v [\phi(z),y]
     \label{nielsenv}
\end{equation}
where
\begin{equation}
     H^v [\phi(z),y]  = -i e^2 v \int d^4x
     \langle \Phi_2(y)\eta(y)\,\bar\eta(x) \Phi_2(x) \rangle_{\rm 1PI} \,\, .
\label{cv1}
\end{equation}
The expansion
\begin{equation}
       H^v[\phi(x),y] = C^v(\phi) + D^v(\phi)(\partial_{\mu}\phi)^2 +\cdots
\label{CDv} 
\end{equation}
starts with the quantity $C^v(\phi)$ which is given
to lowest order by
the diagram in Fig.~5, with the operator
insertions of (\ref{cv1}):
\begin{equation}
C^v_{e^2}(\phi) = - i e^2 \, \int \frac{d^4 k}{(2\pi)^4}
\frac{k^2 - \xi e^2 \phi^2}{D(k) (k^2 + e^2 v \phi)}
\,\,.
\label{Cve2}
\end{equation}
The quantity $D^v (\phi)$ is again
of order one, and proceeding in the same
way as in Section 2, we get
the Nielsen identities for the $v$-dependence
of the effective action:
\begin{equation}
    \frac{\partial V_{e^6}} {\partial v}
    =  C^v_{e^2}\,  \frac{\partial V_{e^4}}{\partial\phi}   \, ,
   \label{Ve6v}
\end{equation}
\begin{equation}
   \frac{1}{2} \frac{\partial Z_{e^2}} {\partial v}
     =   \frac{\partial C^v_{e^2}}{\partial\phi} \, .
   \label{Ze2v}
\end{equation}
Now one can proceed as in (\ref{proof}) to
show that $B_1$ is $v$-independent.

I will now verify the Nielsen identities
(\ref{Ve6v}) and (\ref{Ze2v}).
The first identity can be easily
verified algebraically, without
explicit evaluation of the integrals.
Starting
from (\ref{Ve6}), and using the 
expression (\ref{Ddef}) we get
\begin{eqnarray}
\frac{\partial V}{\partial v} & =&
 - i e^2 \phi  \int \frac{d^4 k}{(2\pi)^4}
\frac{m_2^2(k^2 - \xi e^2 \phi^2)}{D(k) (k^2 + e^2 v \phi)}
\nonumber \\
&=& (\phi m_2^2) \, C^v(\phi) \,\,.
\end{eqnarray}
Using (\ref{m2mass}) this proves
the first identity (\ref{Ve6v}).

The second identity (\ref{Ze2v})
can also be verified algebraically,
but because of the large number of terms
it is easier to evaluate the relevant
expressions explicitly.
The quantity $C^v_{e^2}(\phi)$ is calculated from
(\ref{Cve2}):
\begin{equation}
C^v_{e^2}(\phi) = - \frac{e^2}{(4\pi)^2}
\left[ \ln{\frac{e^2 v \phi}{\mu^2}}
        + \frac{\xi \phi}{2 v} 
        + \frac{1}{2} \right]
\label{eq1}
\end{equation}
where the integral has been 
dimensionally regularized and minimally
subtracted, and $\mu$ is the renormalization scale.

The quantity $Z_{e^2}(\phi)$
can be calculated from the diagrams 
of Fig.~2.
\begin{equation}
       Z = - i \left.{\partial \over \partial p^2} \sum I_j
        \right|_{p^2=0}   \, \, ,
\label{z3}
\end{equation}
where $I_j(p^2)$ is the contribution of the diagram $j$
with two external  $\phi$ lines
with momenta $p$ and $-p$.
We need the contributions of the diagrams that are
proportional to $p^2$ and $v$-dependent.
The sum of the first four is
\begin{eqnarray}
& &I_a + I_b + I_c +I_d =
p^2  e^2 \int \frac{dk^4}{(2\pi)^4} 
\left[ \frac{\xi}{D(k)} +   \right.
\nonumber \\
& & 
\left.
+ \frac{e^2 v^2}{D(k)^2} \left(
\frac{3(e^2 v \phi)^2}{k^2 +e^2 v \phi}
-\frac{3 (e^2 v \phi)^2}{2 k^2}
- 5 e^2 v \phi
-3 k^2 \right) \right] =
\nonumber \\
 & &
= - i p^2 \frac{e^2}{(4 \pi)^2}
\left( \xi \ln{\frac{e^2 v \phi}{\mu^2}}
+ \frac{25 v}{12 \phi} \right)
\,\,.
\label{z1}
\end{eqnarray}
The ghost loop contribution is
\begin{eqnarray}
I_e &=& \frac{1}{2} p^2 e^6 v^3 \phi
\int \frac{d^4 k}{(2\pi)^4}
\frac{1}{(k^2 + e^2 v\phi)^4}
\nonumber \\
&=& i p^2 \frac{e^2}{(4 \pi)^2}
\, \frac{v}{12 \phi}
\,\,.
\label{z2}
\end{eqnarray}

Combining (\ref{z3}), (\ref{z1}), (\ref{z2})
we get for the
$v$-dependent part of $Z(\phi)$:
\begin{equation}
Z_{e^2}^v  = - \frac{e^2}{(4 \pi)^2}
\left(\xi \ln{\frac{e^2 v \phi}{\mu^2}}
    +  2 \frac{v}{\phi}\right)
\,\,.
\label{eq2}
\end{equation}
From the expressions
(\ref{eq1}), (\ref{eq2})
we see that the second Nielsen identity,
(\ref{Ze2v}), is also 
satisfied.

\section{Conlusion}

In this paper I have shown
that the calculation of
the false vacuum decay rate
via a derivative expansion of the 
effective action breaks down
in the case of theories with
radiative symmetry breaking when
the calculation is done in the
class of covariant gauges.

I have also shown that these problems
do not exist in the class of
$R_{\xi}$ gauges, for generic values of the
gauge parameter $v$ that are not
anomalously small.
In this case, the derivative expansion
of the effective action can be used
to calculate the correct, gauge-independent
value of physical quantities.

Considerations of gauge independence
are an important check of the consistency
of the formalism.
I showed that the above results are
reflected in the Nielsen identities
that describe the gauge dependence
of the effective action.
As a final check, I showed that the 
result for the  false vacuum decay rate
in $R_{\xi}$ gauges
is $v$-independent.

It is easy to formally generalize the
calculations of this work and apply it
to other models (non-abelian gauge theories
or high temperature phase transitions)
although the verification of the
Nielsen identities in these cases
may be more involved.
The issue of gauge independence is
still an important one, and considerations
similar to this work may be helpful in order 
to clarify whether various calculations
of physical quantities are self consistent
or plagued by infrared divergences
in different gauges.

\newpage
 
\centering{\epsffile{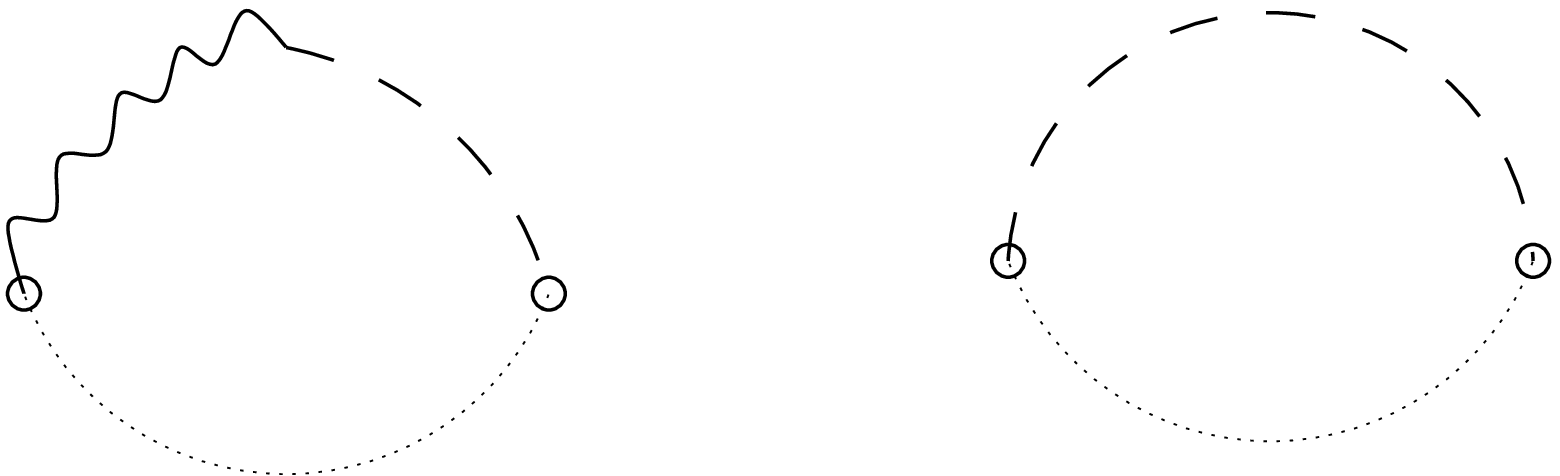}}

\bigskip

Figure 1: The two graphs that contribute to $C_{e^2}$.  Photon,
$\Phi_2$, and ghost propagators are indicated by wiggly, long-dashed,
and short-dashed lines, respectively.
 
\newpage
 
\centering{\epsffile{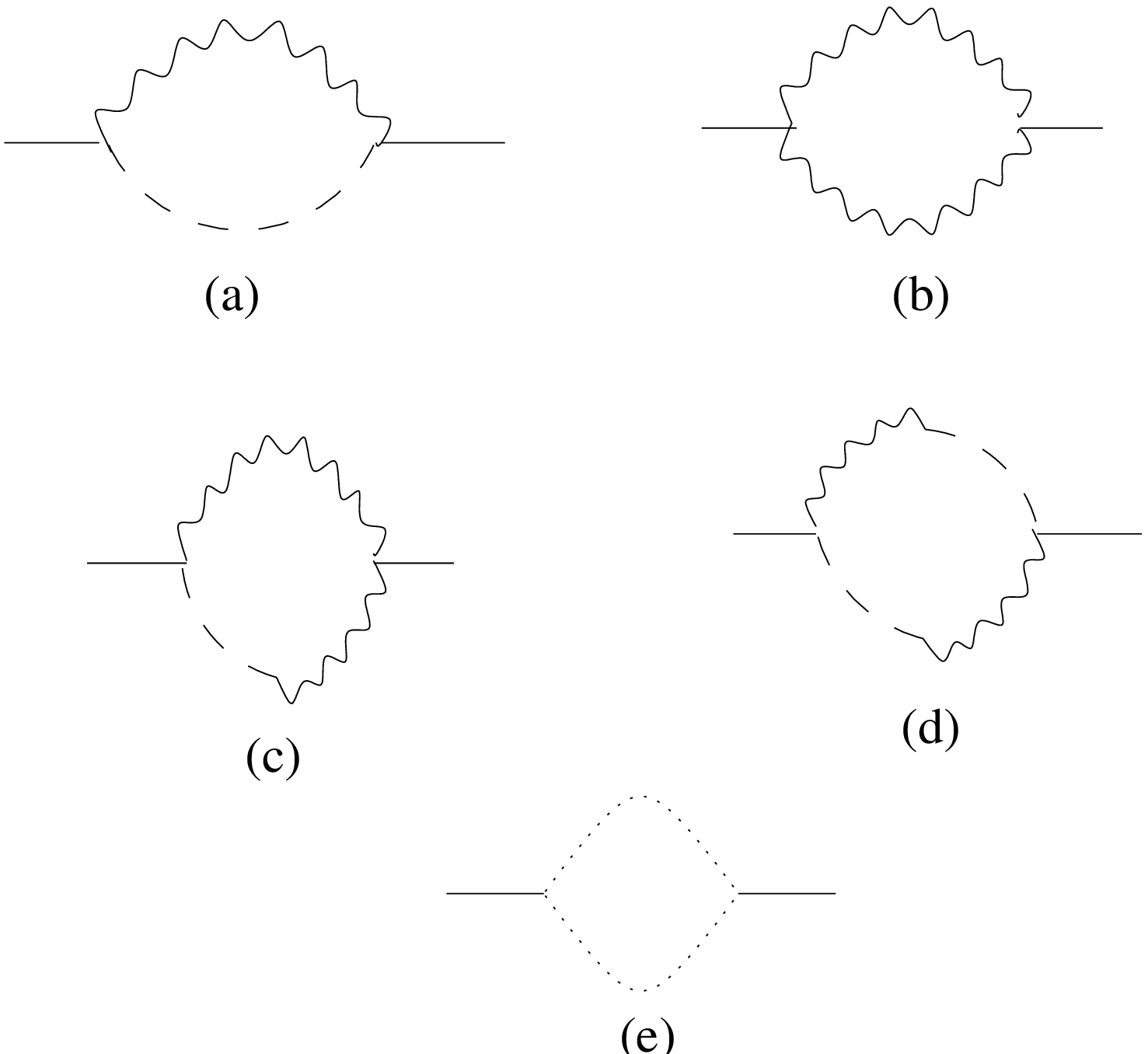}}

\bigskip

Figure 2: The graphs that contribute to $Z_{e^2}$.  

\newpage

\centering{\epsffile{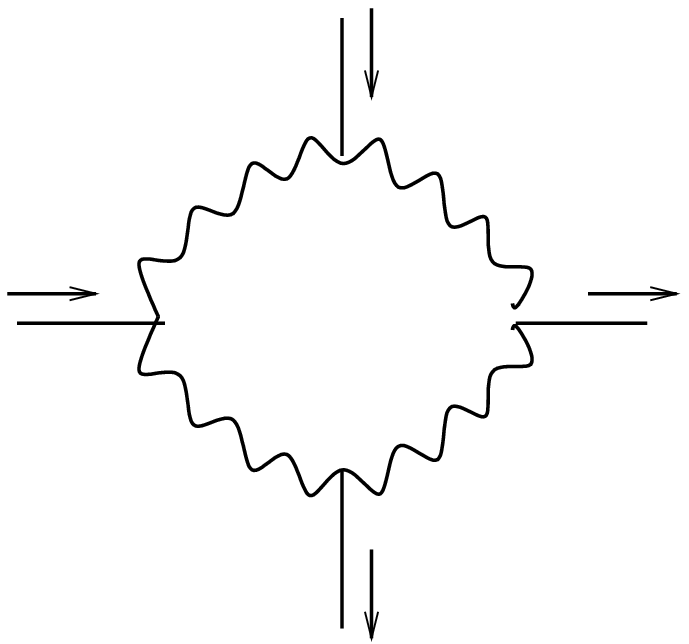}}

\bigskip

Figure 3: Example of a graph that signals
the breakdown of the derivative expansion.

\newpage

\centering{\epsffile{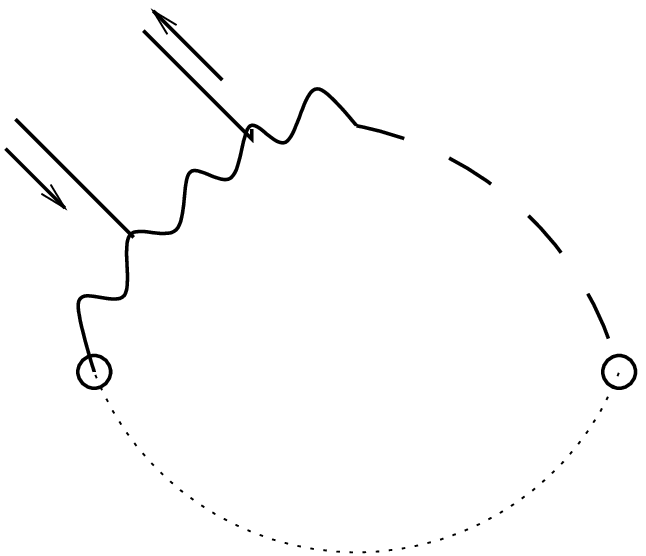}}

\bigskip

Figure 4: Example of a graph that 
contributes to the Nielsen identities.

\newpage

\centering{\epsffile{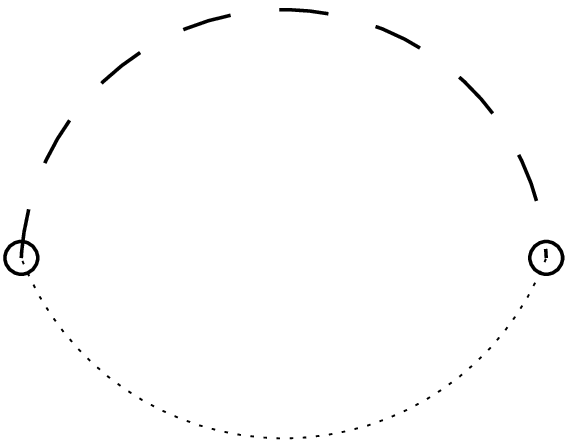}}

\bigskip

Figure 5: The graph that contributes to $C^v_{e^2}$.  

\newpage


\begin{thebibliography}{10}

\bibitem{coleman} S.~Coleman, 
                {\it Phys. Rev.} {\bf D15},
               2929, (1997).
\bibitem{ejw}E.J.~Weinberg, {\it Phys. Rev.} {\bf D47},
                    4614 (1993).
\bibitem{metaxas} D.~Metaxas and E.J.~Weinberg,
         {\it Phys. Rev.} {\bf D53}, 836 (1996).
\bibitem{baacke} J.~Baacke and K.~Heitman,
            {\it Phys. Rev.} {\bf D60}, 105037 (1999).
\bibitem{Kr} J.~Kripfganz, A.~Laser, M.G.~Schmidt,
            {\it Nucl. Phys.} {\bf B433}, 467 (1995).
\bibitem{Bod} D.~Bodeker, W.~Buchmuller, Z.~Fodor, T.~Helbig,
             {\it Nucl. Phys.} {\bf B423}, 171 (1994).
\bibitem{Sur} A.~Surig,
            {\it Phys. Rev.} {\bf D57}, 5049 (1998).
\bibitem{jackiw} R.~Jackiw, 
       {\it Phys. Rev.} {\bf D9}, 1686 (1974).
\bibitem{nielsen} N.K.~Nielsen,
         {\it Nucl. Phys.}
                  {\bf B101}, 173 (1975).  
\bibitem{Kobes}R.~Kobes, G.~Kunstatter and A.~Rebhan, 
         {\it Nucl. Phys.}
                  {\bf B355}, 1 (1991).
\bibitem{stathakis}For an investigation of the issues in these gauges,
   see N.S.~Stathakis, Ph.~D. Thesis, Columbia University (1990).
\bibitem{af}I.J.R.~Aitchison and C.M.~Fraser, {\it Ann. Phys.} {\bf
         156}, 1 (1984).


\end{thebibliography}
\end{document}